\documentclass[prb,showpacs,floatfix,twocolumn]{revtex4}

\usepackage{epsfig,graphicx}

\begin{document}

\title[Magnetic frustration on the diamond lattice]
{Magnetic frustration on the diamond lattice of the A-site magnetic
spinels CoAl$_{2-x}$Ga$_x$O$_4$: The role of lattice expansion and site disorder} 

\author{Brent C. Melot, Katharine Page and Ram Seshadri}
\affiliation{Materials Department and Materials Research Laboratory\\
	 University of California, Santa Barbara CA 93106}

\author{E.M. Stoudenmire and Leon Balents}
\affiliation{Department of Physics,\\
	    University of California, Santa Barbara, CA 93106}

\author{Doron L. Bergman}
\affiliation{Department of Physics, Yale University\\ 
P.O. Box 208120, New Haven, CT 06520-8120}

\author{Thomas Proffen} 
\affiliation{Los Alamos National Laboratory, Lujan Neutron Scattering Center\\ 
LANSCE-12, MS H805, Los Alamos, NM 87545}

\begin{abstract}

The spinels CoB$_2$O$_4$ with magnetic Co$^{2+}$ ions on the diamond lattice
A site can be frustrated because of competing near-neighbor ($J_1$) and 
next-near neighbor ($J_2$) interactions. Here we describe attempts to understand 
these interactions by substitution on the non-magnetic B-site. 
The system we employ is CoAl$_{2-x}$Ga$_x$O$_4$, where Al is 
systematically replaced by the larger Ga, ostensibly on the B site. 
Ga substitution has the effect of expanding the lattice and pushes Co atoms on the A-site  
further away from one another  weakening the magnetic interactions.
We also find, however, that Ga distributes between the B and the A site in a concentration 
dependent manner displacing an increasing amount of Co from the A site for larger values of $x$. 
This site mixing, confirmed by powder neutron diffraction studies carried out at room temperature, 
affects magnetic properties very significantly and changes the nature of the ground state. 
We discuss the role that both structural changes play in changing the degree of magnetic
frustration on the diamond lattice.  
We also use classical Monte-Carlo modeling of the magnetic coupling to illustrate 
the complexity of the interactions that arises from site mixing.

\end{abstract}


\maketitle

\section{Introduction}

The presence of magnetic frustration in the solid state is known to give rise 
to multiple, nearly degenerate and often non-collinear magnetic ground
states.\cite{GeoFrustRev:1994,GreedanFrust:2001,RabeDesign:2006} Interest
in the competition between these ground states has attracted considerable
attention in recent years with the resurgent field of multiferroics: systems
with spiral spin ordering are known to couple spin and lattice degrees of
freedom\cite{Katsura,Mostovoy,Sergienko} and in specific cases, to even give 
rise to spontaneous polarization.\cite{TokuraScience2006,Tokura:2006}

One highly studied class of materials that has been shown to exhibit this 
type of non-collinear ordering resulting from magnetic frustration
is the spinel family.\cite{CCO:2006,MnO:2007} 
Spinels have the general formula AB$_2$X$_4$, and in so-called normal spinels,
the A site is divalent and tetrahedrally coordinated by the anion X, while the 
B site is trivalent and octahedrally coordinated by the anion X (typically O 
or S). Both sites can accommodate magnetic cations allowing for a wide range of 
magnetic properties. It is important to note that these sites are often not 
well ordered and as a consequence it is frequently appropriate to write the 
formula as (A$_{1-\delta}$B$_{\delta}$)(A$_{\delta}$B$_{2-\delta}$)X$_4$ 
where $\delta$ is referred to as the inversion parameter and can be any 
value 0 $\le \delta \le$ 1. 

\begin{figure}
\centering\includegraphics[width=7cm]{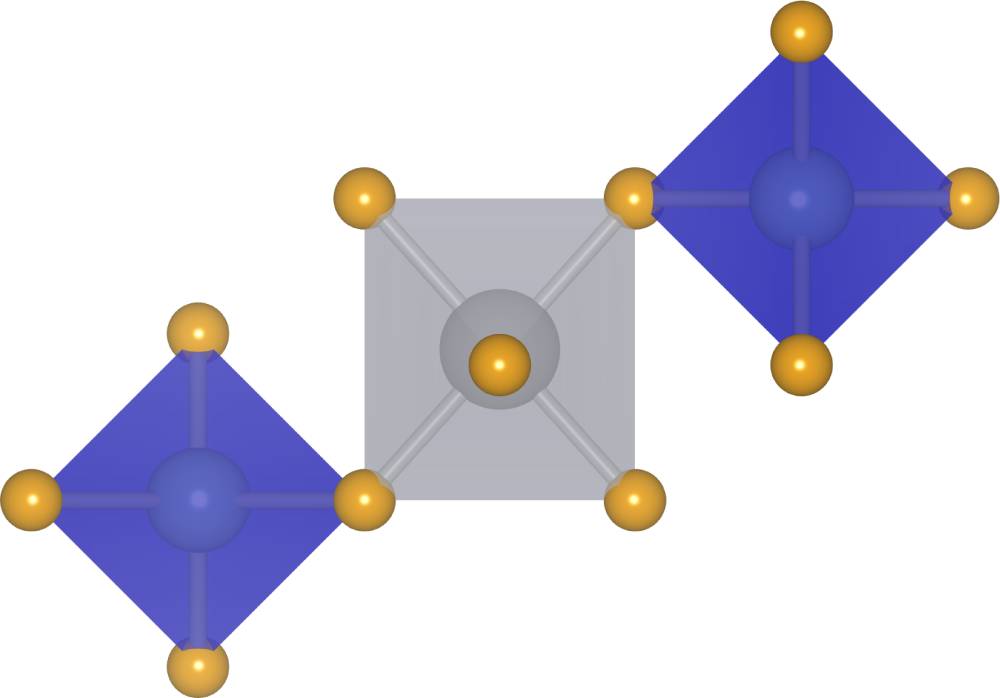}\\
\caption{(Color online) Illustration of the superexchange pathway for magnetic 
interactions between neighboring A-site atoms in the AB$_2$O$_4$ spinel structure. 
A atoms are blue (dark grey), B are light grey (light greay) and O are the orange (grey)
spheres.  Note that the near-neighbor $J_1$ coupling proceeds from each A atom to 
one of it's 4 near-neighbor A atoms through this 6-fold degenerate pathway. 
The next-near-neighbor $J_2$ coupling to one the 12 next-near neighbors is doubly 
degenerate and proceeds through an identical pathway.}
\label{fig:struc}
\end{figure}

The spinel structure can be viewed as two interpenetrating sublattices 
with B atoms forming a pyrochlore lattice while the A atoms constitute a 
diamond lattice.\cite{BsitePyrochlore:1956} It is well known that there are 
multiple magnetic interactions in these systems. Here we focus mostly on the 
nearest neighbor ($J_1$) and next nearest neighbor ($J_2$) exchange 
interactions on the diamond lattice using the compound CoAl$_2$O$_4$ as a 
starting point (Fig.\,\ref{fig:struc}). Since the diamond lattice can be 
broken down further into two interpenetrating face-centered cubic (\textit{fcc})
sublattices, we can envisage $J_1$ as coupling the two \textit{fcc} sublattices
to each other and $J_2$ coupling nearest neighbors within each 
sublattice.\cite{BalentsOrderbyDisorder} 

In a recent work Tristan \textit{et al.} demonstrated that normal spinels, 
$\delta \ll 1$, with magnetic cations on the A site and non-magnetic cations 
on the B site can exhibit strong magnetic frustration.\cite{Tristan:2005}
These A-site magnetic spinels have been studied in the past by
Roth\cite{RothJPhysFrance} and Blasse.\cite{BlasseNewSuperExchange}
Recent modelling studies by Bergman \textit{et al.} suggested that the 
magnitude of the ratio $J_2/J_1$ can strongly influence the dominant magnetic
ground state in these systems.\cite{BalentsOrderbyDisorder} In the limit that
$J_2/J_1 \approx 0$, the magnetic ordering is a simple N\'eel state
(magnetically speaking, a diamond to zinc blende transition). As the ratio
becomes larger, the nature of the magnetic ground state can change from
N\'eel to a complex spiral pattern with the spirals oriented parallel to
(111) planes. Tristan \textit{et al.}\cite{Tristan2008} have also recently examined 
solid solutions between CoAl$_2$O$_4$ and Co$_3$O$_4$ to investigate the effect of the 
B site cation in the superexchange pathways that connects neighboring A sites.

These recent studies have encouraged us to attempt to alter this ratio of 
$J_2/J_1$ in these A-site magnetic spinels by controlling interatomic 
spacings through appropriate substitution on the B site of the spinel 
structure. We use Ga substitution ostensibly on the B site of spinels 
CoAl$_{2-x}$Ga$_x$O$_4$ as a means of separating the Co from one another, 
given that Ga$^{3+}$ is significantly larger than Al$^{3+}$ (the respective 
6-coordinate radii are 0.62\,\AA\/ and 0.535\,\AA).\cite{Shannon} We find that 
while the effect of such substitution is indeed to steadily increase the distance 
between neighboring Co atoms, there is the additional effect that site inversion 
in the structure steadily increases simultaneously. While separating the effects 
of these simultaneous changes in the structure is dificult, we discuss the role
each change plays in the context of the altered frustration parameter.

\section{Experimental details}

Polycrystalline samples of the compounds were prepared using ceramic routes.  
Cobalt oxalate (CoC$_2$O$_4\cdot$2H$_2$O) was mixed with stoichiometric 
amounts of Ga$_2$O$_3$ and Al$_2$O$_3$ and intimately ground with ethanol
in an agate mortar. The powders were then pressed into 13\,mm pellets
and fired in air in alumina crucibles at 800$^\circ$C for 24\,h. The
pellets were then reground, pressed again into pellets, and fired at
1200$^\circ$C (1000$^\circ$C for $x$=0.0) for 12\,h.  In order to 
obtain equilibrated samples, all pellets were annealed by heating to 
700$^\circ$C for 12\,h, cooling to 400$^\circ$C at a rate of 
3$^\circ$C\,min$^{-1}$, soaking at 400$^\circ$C for 
120\,h\cite{CGOcatdistro:1980} followed by cooling in the furnace to the room 
temperature. For all heat treatments, pellets were placed on a
bed of powder with the same stoichiometry to minimize reaction with the
crucible.

X-ray diffraction patterns were obtained using CuK$\alpha$ radiation on a
Philips XPERT MPD diffractometer operated at 45\,kV and 40\,mA.  Phase purity
was determined by refining the patterns using the Rietveld method
as implemented in the \textsc{xnd} Rietveld code.\cite{XND}   Neutron 
diffraction data were collected on the neutron powder diffractometer (NPDF) 
at the Lujan Center at Los Alamos National Laboratory at room temperature,
on samples sealed in vanadium cans.\cite{NPDF} Neutron diffraction data were 
refined using the Rietveld method as implemented in the \textsc{expgui-gsas} 
software suite.\cite{gsas,expgui}. Local structures as obtained from pair 
distribution function (PDF) analysis of the total neutron scattering were extracted 
from the total scattering data using the program PDFgetN\cite{PDFGETN} and 
with a maximum momentum transfer $Q_{max}=40$\,\AA$^{-1}$. The obtained PDFs were analyzed using the \textsc{PDFgui} software package.\cite{diffpy} 
DC magnetization was measured using a Quantum Design MPMS 5XL SQUID 
magnetometer.

\section{Computational Details}

Classical Monte Carlo simulations of the magnetic behavior of the system
were performed using the ALPS project's spinmc application.\cite{ALPS} 
A custom lattice with periodic boundary conditions was generated for each simulation run in order to allow disorder
averaging. The lattice generation code first randomly selected an A site 
and then moved its spin to a B site, also chosen at random and independent of the position
of the A site. The simulation was then run using a Heisenberg 
Hamiltonian with $J_1$ and $J_2$ bonds as described above, 
and with $J_i$ impurity bonds connecting
spins on B sites with their nearest-neighbor occupied A sites. 
Finally, because the typical error bars obtained from each Monte Carlo susceptibility 
simulation were negligible (about $1 \times 10^{-4}$ relative error), 
the error bars plotted represent only variation due to the presence of disorder as obtained by 
averaging over several runs.

A numerically calculated inverse susceptibility curve was fit to the experimental curves as 
follows: first, using the fact that the Curie-Weiss temperature for the $J_1$-$J_2$-$J_i$ model 
with an inversion parameter, $\delta$, is given by

\begin{widetext}

\begin{equation}
\Theta_{\rm{CW}}/J_1 = -\frac{4 S (S+1)}{3}\left[(1+3J_2/J_1)(1-\delta)^2 + 3J_i/J_1 (\delta-\delta^2) \right],
\label{eqn:theta_cw}
\end{equation}

\end{widetext}

\noindent
one can calculate the value of $J_1$ necessary to match the experimentally measured Curie-Weiss temperature.
This value of $J_1$ is then used to scale the temperature by a factor of $J_1\,S(S+1)$ and the inverse
susceptibility by a factor of $(k_B\,J_1\,S(S+1))/(\mu_{\rm{eff}}^2 \mu_B^2 N_A)$.
Such a fitting procedure always guarantees that the high temperature behavior of the simulation data is in
exact agreement with that of the experimental data.

\section{Results} 

\subsection{Structure} 

\begin{figure} 
\centering \includegraphics[width=8cm]{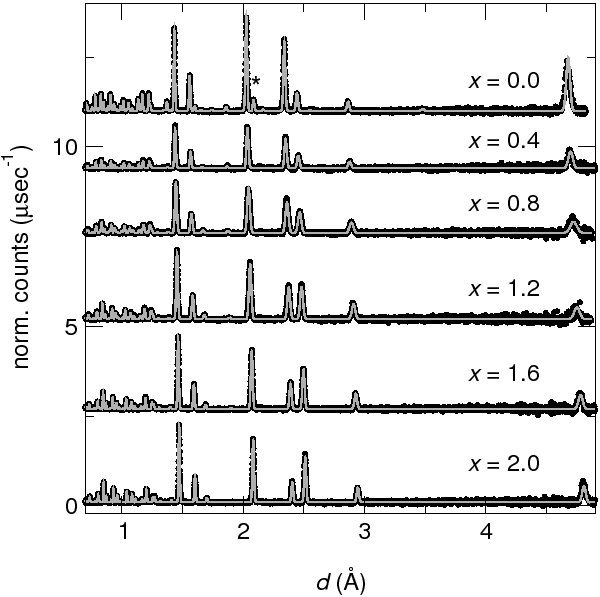}
\caption{Highest $d$-spacing bank of powder neutron diffraction data from the
different samples of CoAl$_{2-x}$Ga$_x$O$_4$ collected at room tempearture. 
Circles are data and solid lines are Rietveld fits to the spinel phase 
[space group $Fd\bar3m$, A $(\frac18,\frac18,\frac18)$, B $(\frac12,\frac12,\frac12)$, 
O $(u,u,u)$ ]. 
Asterisks denote locations where contributions to the scattering from the 
Al$_2$O$_3$ impurity arise.}
\label{fig:neutrons}
\end{figure}

\begin{figure}
\centering \includegraphics[width=8cm]{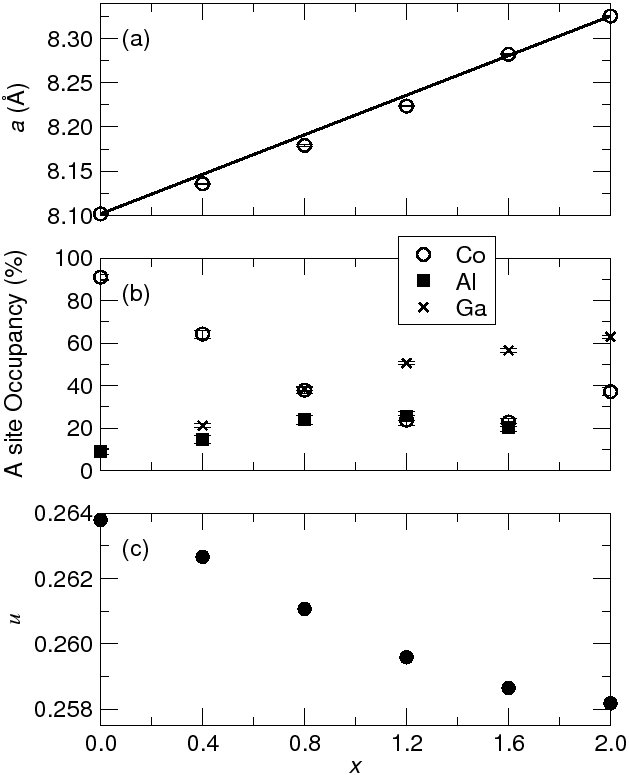}
\caption{Evolution of different refined structural parameters with $x$ for the
different CoAl$_{2-x}$Ga$_x$O$_4$ samples obtained from time-of-flight neutron 
diffraction. (a) The cell parameter, showing a nearly increase in size. The line 
connects end members to illustrate the V\'egard law. (b) Occupancies 
of the different cations on the A site. (c) Internal parameter $u$ 
indicative of the oxygen position. $u = \frac14$ means the BO$_6$ 
octahedra are perfectly regular.}
\label{fig:cell_occ_u}
\end{figure}

Rietveld analysis of the highest $d$-spacing bank (from four banks 
of data for each sample) of time-of-flight neutron diffraction data
are displayed in Fig.\,\ref{fig:neutrons} for the different spinel samples.
There were no peaks in the diffraction data that could not be assigned
to either the spinel phase or a small Al$_2$O$_3$ impurity in the $x$ = 0.0 sample,
 indicating no magnetic impurities and that a complete solid solution is achieved 
for all values of $x$. The oxygen stoichiometry refined within error
to the  correct stoichiometric value excluding the possibility that any    
of the Co$^{2+}$ was oxidized to Co$^{3+}$.  Given the close-packed nature
of the oxygen lattice this was not surprising.  From the Rietveld analysis, evolution of the lattice parameter, 
site occupancy of the cations on the A site, and the internal structural 
parameter (the $u$ of oxygen) are presented in the different panels of
Fig.\,\ref{fig:cell_occ_u}. It is seen that substitution of the 
larger Ga$^{3+}$ for Al$^{3+}$ results in the spinel unit cell edge increasing from 8.1\,\AA\/
to 8.3\,\AA. The V\'egard law is not strictly followed, and for all 
intermediate $x$ values, the cell parameter is slightly reduced from the
values suggested by a weighted average of the end-members.

Panel (b) of Fig.\,\ref{fig:cell_occ_u} shows the results of allowing all ions,
Al, Ga, and Co to distribute themselves between the A and B sites in 
the refinements, with the constraint that the total amounts of the different 
atoms were as dictated by the starting stoichiometry. The process of achieving
this refinement involved declaring on each of the A and B sites, two separate 
Co atoms, one of which exchanged with Al and the other with Ga. A related
procedure has been described by Joubert \textit{et al.}\cite{Joubert} 
Due to the nature of multiple constraints in the refinements, errors on 
the occupancies are likely to be underestimated. It is noted from 
Fig.\,\ref{fig:cell_occ_u}(b) that as larger amounts of Ga are substituted into 
the system an increasing displacement of Co from the tetrahedral site to the 
octahedral site occurs. While it might be expected that the Ga$^{3+}$ ions 
would prefer the octahedral site based upon its large size, the amount of 
Ga$^{3+}$ found on the smaller tetrahedral site steadily increases across 
the substitution series. In contrast Al$^{3+}$ ions are found to remain mostly 
normal with a tetrahedral site occupancy never exceeding 20\,\%. 
The site preference of Ga$^{3+}$ for the tetrahedral site is in agreement 
with the tendency of $d^{10}$ Ga$^{3+}$ to adopt $sp^3$ hybridization. 
In fact, from electrostatic arguments Miller\cite{Miller1959} has determined 
that the relative octahedral site preference of Al$^{3+}$ and Ga$^{3+}$ are respectively 
$-$2.5\,kcal\,mol$^{-1}$ and $-$15.4\,kcal\,mol$^{-1}$ with the larger number
indicative of the greater octahedral preference. In contrast, Co$^{2+}$
has a site preference energy of $-1$10.5\,kcal\,mol$^{-1}$.
Nakatsuka \textit{et al.}\cite{nakatsuka_cation_2003} have conducted a 
recent study on the energetics of different local bonding configurations and 
found that replacing the relatively large Co$^{2+}$ ($r_{tet}$ = 0.58\,\AA\/) 
on the tetrahedral site with Al$^{3+}$ ($r_{tet}$ = 0.39\,\AA\/)
results in abnormally long bond lengths which is not favored.
This effect is not as pronounced when Ga$^{3+}$ moves to the tetrahedral site
given the larger radius ($r_{tet}$ = 0.47\,\AA\/) and that the end member 
CoGa$_2$O$_4$ is commonly found to be almost completely 
inverted.\cite{Nakatsuka:wm2007} Site mixing also serves to explain the 
deviation from the V\'egard law that is observed in Fig.\,\ref{fig:cell_occ_u}(a). 

In Fig.\,\ref{fig:cell_occ_u}(c), the internal parameter $u$ reflecting the
position of oxygen is displayed for the different compounds in the series.
It is seen that with increasing $x$, this value progressively decreases.
Hill \textit{et al.}\cite{Hill1979} have pointed out that in normal spinels
(without any inversion) this parameter $u$ depends on the ratio $R$ of the 
octahedral to tetrahedral bond lengths according to:

\begin{equation}
u = \frac{R^2/4 -2/3 + (11R^2/48 - 1/18)^{1/2}}{2R^2 - 2}
\label{eqn:u}
\end{equation}

\noindent The expected values for ordered CoAl$_2$O$_4$ and CoGa$_2$O$_4$
using the appropriate ionic radii\cite{Shannon} would be respectively 0.265 
and 0.261. The observed trend of a decreasing $u$ agrees with this expectation 
for the normal end-members. However, we see from Fig.\,\ref{fig:cell_occ_u}(c) 
that while CoAl$_2$O$_4$, with very small inversion
has an experimentally determined $u$ value very close to what is calculated 
from equation\,\ref{eqn:u}, the value determined for CoGa$_2$O$_4$ 
($u$ = 0.258) is significantly smaller than suggested by 
equation\,\ref{eqn:u}.  Again, we believe this discrepancy arises because 
of the growing inversion in the compounds as $x$ increases.

\begin{figure}
\centering \includegraphics[width=8cm]{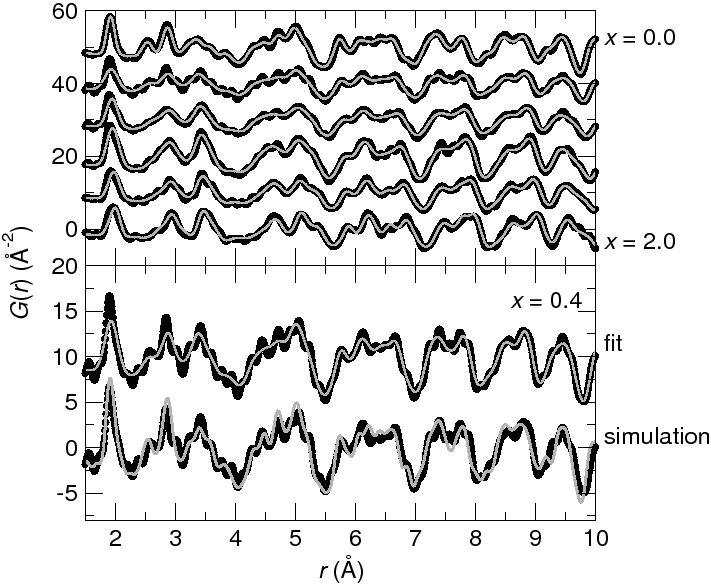}
\caption{(a) Neutron pair distribution functions (PDF) collected at
room temperature for the different 
CoAl$_{2-x}$Ga$_x$O$_4$ samples with $x$ = 0.0 at the top of the panel and 
$x$ = 2.0 at the bottom.  Experimental data are filled grey circles and the 
lines are fits obtained by refining data against the average spinel
structure. (b) Experimental PDFs of the $x$ = 0.4 sample (filled grey circles)
compared with the refinement using the average structure (labeled ``fit'') and
compared with a simulation (labeled ``simulation'') prepared by taking the
weighted average of experimental PDFs of the $x$ = 0.0 sample (80\,\% 
weighting) and $x$ = 2.0 (20\,\% weighting).} 
\label{fig:pdf}
\end{figure}

The pair distribution functions extracted from total scattering 
neutron diffraction are shown for each composition in Fig.\,\ref{fig:pdf}(a)
along with fits (shown as lines) obtained by refining the average spinel unit 
cells with the appropriate site inversion. Several characteristics can be 
observed directly. The unit cell expansion is apparent in the movement of 
atom-atom peaks to longer $r$ with increasing $x$. The first atom-atom peak 
become increasingly broad as $x$ increases, perhaps as a result of 
increased site mixing, in turn resulting in a distribution of cation-oxygen 
distances. Furthermore, features in the intermediate compositions can be 
linked to the CoAl$_2$O$_4$ and CoGa$_2$O$_4$ end-member PDFs, with an 
increasing resemblance to the CoGa$_2$O$_4$ PDF as $x$ increases. 

Refinements in real space were carried out starting with the average structure
results.  A stable refinement of the site occupancies could not be achieved 
and values were therefore fixed to the results obtained from the Rietveld 
analysis of the neutron scattering data. The parameters refined in the PDF 
analysis include lattice parameters, isotropic atomic displacement parameters 
for each atomic species, a scale factor, and quadratic peak sharpening. The 
refinements yielded $\chi^2 < 1$ and $R_{wp}$ less than  15\,\% for 
$r$-ranges from 1 to 20\,\AA. From the fits of the PDF displayed in 
Fig.\,\ref{fig:pdf}(a), it is seen that of all the samples, only in the data 
corresponding to the $x$ = 0.0 sample are all the peaks well described by the 
model. This is keeping with the fact that only in the $x$ = 0.0 sample does
the average spinel structure accurately describe all the local distances, 
since this structure has the lowest inversion and no substitution.

In Fig.\,\ref{fig:pdf}(a), this is shown for a particular sample with 
$x$ = 0.4. The experimental PDF is only very poorly fit in the short $r$ 
region by the average structure model (labeled ``fit''). Upon close 
inspection of the $x$ = 0.4 refinement around 3\,\AA\/ and 5\,\AA\/, which 
correspond respectively to the first and second nearest Co neighbors, 
there is significant splitting of the experimental peaks which is not well 
represented by the average structure. However, a much better description of
the $x$ = 0.4 sample is a stoichiometrically weighted (80:20) average of the 
experimental PDFs of the end members ($x$ = 0.0 and $x$ = 2.0) rather than 
as a single phase as illustrated in Fig.\,\ref{fig:pdf}(b). In this, a 
resemblance to systems such as In$_{1-x}$Ga$_x$As\cite{JeongPRB} is noted
wherein the alloy compositions locally follow the bonding rules of the 
end-member structures, and alloy PDFs can be described using weighted averages 
of the end members. This result of averaging end-members only holds true for 
distances within a unit cell (approximately 8\,\AA). Outside of this range the 
superposition model begins to fail and the average structure proves to be a 
good model of the data.  Thus from the local structure analysis we can 
see that our samples have a homogeneous distribution of substituted cations
and there does not seem to be any evidence for local clustering of cations.  

\subsection{Magnetism} 

\begin{table*}[t]
\caption{Magnetic data of different spinel compounds. $\Theta_{CW}$ is obtained
from the fit to the high temperature inverse suscpetibility data as described in 
the text. $T_N$ is taken as the point where
the ZFC and FC magnetization curves diverge with respect to temperature. 
  $f$ is obtained from $\Theta_{CW}/T_N$.
The inversion parameter $\delta$ is also indicated for all samples.}
\label{tab:magn}
\begin{tabular}{lcccccc}
\hline \hline 
Compound       & $\delta$ & $\mu_{eff}$ ($\mu_B$) & $\Theta_{CW}$ (K) & $T_N$ (K) & $f$ & Reference\\
\hline \hline
$x$ = 0.0      &  0.09     & 4.59         & -103    & 12   &  8.6 & this work \\
$x$ = 0.4      &  0.36     & 4.80         & -96     & 5.3  &  18  &           \\
$x$ = 0.8      &  0.62     & 4.83         & -75     & 7.3  &  10  &           \\
$x$ = 1.2      &  0.76     & 4.83         & -54     & 8.3  &  6.5 &           \\
$x$ = 1.6      &  0.77     & 4.80         & -45     & 9.0  &  5.0 &           \\
$x$ = 2.0      &  0.63     & 4.84         & -42     & 9.2  &  4.6 &           \\
\hline
CoGa$_2$O$_4$  &  0.29     & 4.96         & -55    & 10 & 5.5 & \onlinecite{Fiorani1978}\\
\hline
CoAl$_2$O$_4$  &  0.04     &              & -89    & 9 & 10 & \onlinecite{SuzukiCoAl2O4}\\
\hline
Co$_3$O$_4$    &  0.00     &              & -110   & 30 & 3.7 & \onlinecite{SuzukiCoAl2O4}\\
\hline
CoRh$_2$O$_4$  &  0.00     &              & -31    & 25 & 1.2 & \onlinecite{SuzukiCoAl2O4}\\
\hline \hline
\end{tabular}
\end{table*}

\begin{figure}[b]
\centering \includegraphics[width=8cm]{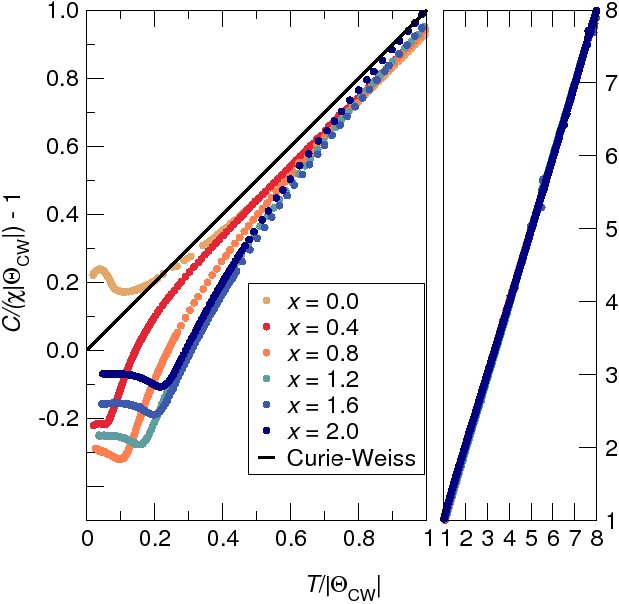}
\caption{(Color online) Temperature dependence of the scaled zero-field cooled magnetic 
susceptibility of the different spinel samples. Data were acquired under a 
100\,Oe field. The nature of the scaling is described in the text.}
\label{fig:curie}
\end{figure}

Figure\,\ref{fig:curie} displays the temperature dependence of the DC magnetic
susceptibility $\chi$ of the different spinel samples on a single scaled plot.
For all samples, data between 350\,K and 400\,K were fit with high reliability
by the Curie-Weiss formula: $\chi = C/(T-\Theta_{CW})$, where $C$ is the 
Curie constant, and $\Theta_{CW}$ is the Curie-Weiss ordering temperature.
We can recast this formula as: 

\begin{equation}
\frac{C}{\chi|\Theta_{CW}|} = \frac{T}{|\Theta_{CW}|} - 1
\label{eqn:curie}
\end{equation}

\noindent Evidence of Curie-Weiss behavior at high temperatures is seen from
the right, higher temperature, panel of Fig.\,\ref{fig:curie}. For each sample,
and at all temperatures above the individual $\Theta_{CW}$ (indicated in
Table\,\ref{tab:magn}, and approximately ranging between -100\,K and -40\,K)
the scaled inverse susceptibility (left-hand side of equation\,\ref{eqn:curie}) is 
precisely equal to $T/|\Theta_{CW}|$. All the samples order at temperatures
well below $|\Theta_{CW}|$ as seen in the left-hand low-temperature panel of Fig.\,\ref{fig:curie}. 
A curious point to note is that while all the samples display a downturn in 
plots of $\chi$ \textit{vs.\/} $T$, it is only the $x$ = 0.0 sample,
CoAl$_2$O$_4$, that stays largely superior to the dotted Curie-Weiss line. The behavior of 
$x$ = 0.0 is distinctly different from all other samples which progressively 
deviate from the dotted Curie-Weiss line at higher temperatures
as $x$ increases. The utility of plotting the inverse susceptibility
of a solid solution in the manner shown in Fig.\,\ref{fig:curie} becomes 
evident in the way compounds with uncompensated spins due to inversion 
($x \ge 0.4$) are separated from the antiferromagnetic $x$ = 0.0 end member~\cite{Melot2009}. 
A plot of $1/\chi$ \textit{vs.} $T$ does not reveal this since the magnititude 
of the susceptilities varies through the series. 

Table\,\ref{tab:magn} also shows values of $\mu_{eff}$ obtained from the
Curie constant $C$ for the different spinel samples. The expected spin-only 
value of the magnetic moment for tetrahedral Co$^{2+}$ is  3.88\,$\mu_B$
whereas a value of 5.20\,$\mu_B$\cite{DaySelbin} is expected for systems
with completely unquenched orbital contribution. The values
obtained here run between 4.59\,$\mu_B$ and 4.85\,$\mu_B$, and are therefore
sandwiched by the limits of the completely quenched and unquenched orbital 
contributions. Measured values from the literature are in the range of 
4.4\,$\mu_B$ to 4.8\,$\mu_B$.\cite{DaySelbin}
Cossee and van Arkel\cite{Cossee_JPhysChemSolids1960} have argued that
for tetrahedral Co$^{2+}$, the proximity of a low-lying excited spin state
adds a temperature-independent term to the Curie-Weiss law and that after
making such a correction, magnetic moment is close to 4.4\,$\mu_B$. 
We expect CoAl$_2$O$_4$, with a very small degree of inversion, to display
a $\mu_{eff}$ value which is close to 4.4\,$\mu_B$. We find that 
for $x \ne 0$ in the substitution series, the value of $\mu_{eff}$ 
is always larger than for $x$ = 0.0, which we attribute to the increasing 
amount of octahedral Co$^{2+}$ which has a larger orbital contribution to the 
effective moment and correspondingly has an experimental range of $\mu_{eff}$ 
from 4.7\,$\mu_B$ to 5.2\,$\mu_B$.\cite{DaySelbin}

\begin{figure}
\centering \includegraphics[width=8cm]{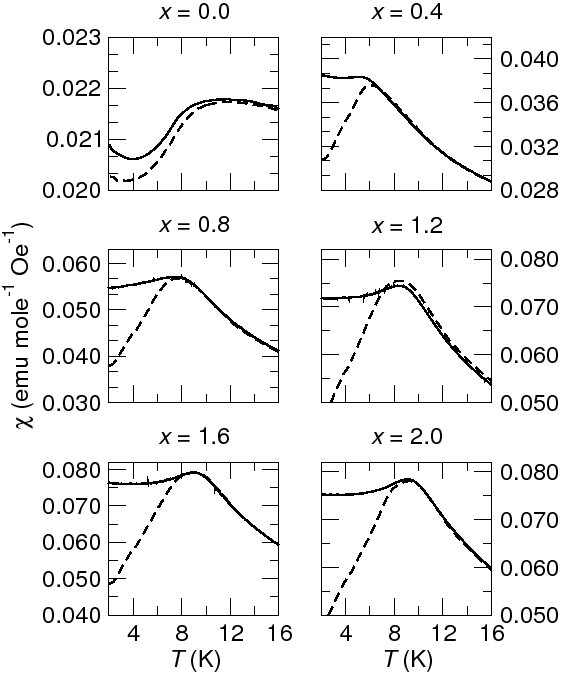}
\caption{Field-cooled (FC, solid) and zero-field cooled (ZFC, dashed) 
molar DC susceptibility of the spinel compounds at low temperatures. Data
were acquired under a magnetic field of 1000\,Oe.}
\label{fig:transitions}
\end{figure}

Figure\,\ref{fig:transitions} shows in closer detail, the temperature dependence 
of the magnetic susceptibilities of the title spinel compounds at low temperatures. All the
compounds display splitting of the FC and ZFC data between 4\,K and 12\,K. CoAl$_2$O$_4$
($x$ = 0.0) shows very little irreversibility and only a gentle downturn 
near 12\,K.  Below 4\,K, there is a small upturn in the
susceptibility which could arise from uncompensated spins. All other 
compounds with $x \ge 0.4$ show the characteristic cusps of glassy systems 
associated with freezing of spins and no long range order as may be expected for crystographically
disordered antiferromagnets.  For all samples $T_N$ was taken as the point of splitting between 
the ZFC and FC curves.

\begin{figure}
\centering \includegraphics[width=7.5cm]{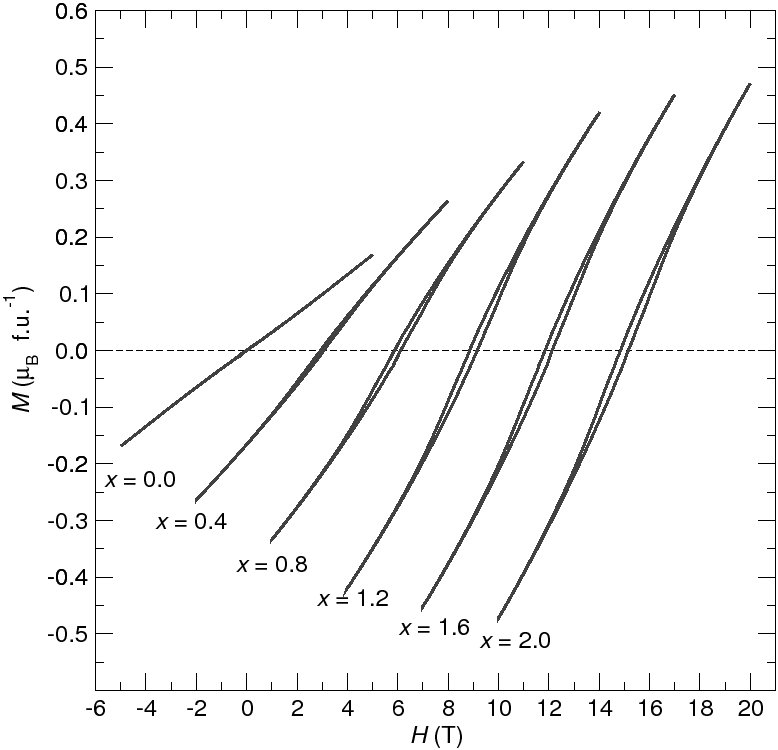}
\caption{Magnetization as a function of magnetic field for all samples 
obtained at 2\,K in fields up to 5\,T.  Note that each curve has been shifted in 3\,T increments
to show more clearly the opening of the loops with increasing $x$.}
\label{fig:hys}
\end{figure}

A gradual opening of the $M-H$ traces as $x$ is increased is seen in 
Fig.\,\ref{fig:hys} which accompanies the increasing concentration of 
Co atoms on the octahedral site.
As the magnetic ions enter the B site the A-B interaction begins to dominate~\cite{BlasseNewSuperExchange, ederer:064409},
and even though both sites have the same number of spins, this interaction can give rise 
to uncompensated spins which could open the $M-H$ loops due to the unequal number of A and B sites.
The near neighbor antiferromagnetic interaction between Co on the A and B sites,
which are only separated by a single O atom will compete strongly with the 
pure A-A interactions where the magnetic ions are separated by $-$O$-$B$-$O$-$ linkages.
This interaction thereby adds another competing exchange pathways which may prevent 
long range antiferromagnetic order between Co atoms on the tetrahderal sites from 
being achieved and give rise to glassy behavior instead.\cite{SpinGlassCGO:1986} 
It should also be noted for high enough concentrations of Co$^{2+}$ on the B site that a 
ferromagnetic direct exchange between neighboring B-B atoms will begin to arise 
which could also give rise to the open $M-H$ loops.~\cite{Diaz2006}

\begin{figure}
\centering \includegraphics[width=8.5cm]{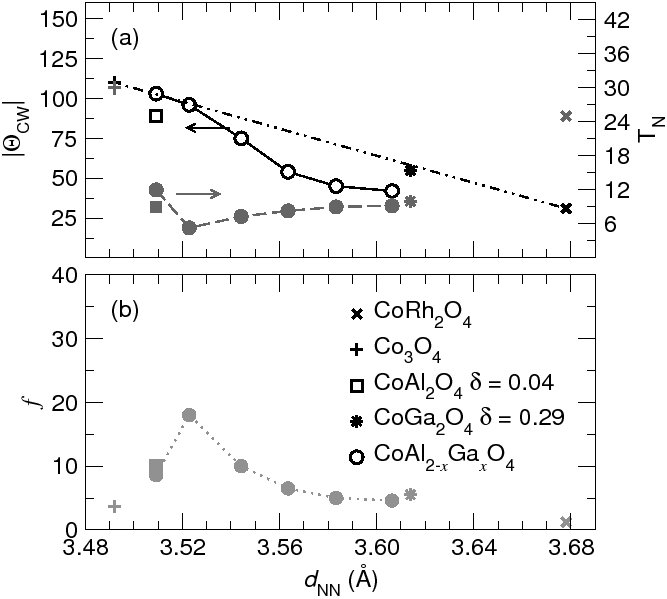}
\caption{(a) Curie-Weiss theta ($\Theta_{CW}$) and ordering temperature ($T_N$)
as a function of near-neighbor spacing between the A site atoms.  The ordering 
temperature is taken as the point of deviation between the field cooling and 
zero field cooling data shown in figure \ref{fig:transitions}.
(b) Frustration index $f = \Theta_{CW}/T_N$ for the title compounds as a
function of near-neighbor spacing between the A site atoms ($d_{AA}$ increases with
increasing Ga content, $x$).  
Also displayed is published data for CoAl$_2$O$_4$ with a different degrees of 
inversion: $\delta$ = 0.04. By comparison, the degree of inversion 
of the CoAl$_2$O$_4$ sample measured in this work is close to 0.09.  
Values for Co$_3$O$_4$ and CoRh$_2$O$_4$ were also obtained from reference 
\onlinecite{SuzukiCoAl2O4}  while values for CoGa$_2$O$_4$ were taken from 
\onlinecite{Fiorani1978}.  Note that the values from  reference 
\onlinecite{SuzukiCoAl2O4} were not reported with lattice constants 
so the near-neighbor separation was taken from separate structure reports.
}
\label{fig:frust}
\end{figure}

For higher concentrations of Ga, it can be seen in Table \ref{tab:magn} and Fig. \ref{fig:frust} (a) 
that $\Theta_{CW}$ gradually decreases. Comparing the title compounds with systems that have a 
well ordered magnetic A lattice such as Co$_3$O$_4$ and CoRh$_2$O$_4$ shows
an almost linear dependence of $\Theta_{CW}$ with respect to nearest neighbor separation $d_{AA}$
(and correspondingly next-near neighbor A-A separation). 
We note that the deviation from the linear trend for samples with $x\ge0.8$ can 
be understood by the site mixing in the samples 
reducing the number of magnetic cations on the A site.  Further evidence that this 
the deviation from linearity is a result of site disorder is the fact that a separate 
report on a sample CoGa$_2$O$_4$ with half of the site mixing present in our samples
lays on the line.  

The structural changes also affect the temperature where
the system transitions from the paramagnetic to glassy state which we take
to occur where the zero field cooling and field cooling data deviate as discussed earlier 
(Fig. \ref{fig:frust} (a)). For small concentrations of Ga with $x$ = 0.4, a sharp drop of the transition
temperature to 5\,K from 10\,K for the pure CoAl$_2$O$_4$ which we 
attribute to the sudden increase in atomic disorder and dilution of the magnetic A site lattice~\cite{Binder1986}.  
This decrease is then followed by a gradual increase as more Al is replaced by Ga.  
Tristan {\it et al.} have studied the effect of replacing the Al$^{3+}$ with
non-magnetic octahedral Co$^{3+}$.~\cite{Tristan2008} It is interesting to note that changes in the
non-magnetic B site cation result in little to no change in $\Theta_{CW}$, however a clear increase
in the ordering temperature is observed.

Figure\,\ref{fig:frust} (b) plots the measured frustration parameter, 
$f = \Theta_{CW}/T_N$, as a function of $d_{AA}$, the separation between 
A ions in the different structures, including values for Co$_3$O$_4$ and 
CoRh$_2$O$_4$ taken from the literature. 
Considering the changes in $\Theta_{CW}$ and $T_N$ we find with the exception of $x$ = 0.4
the frustration index decreases systematically with increasing $d_{AA}$ separation.  

In order to better understand the behavior of the samples, we performed
classical Monte Carlo simulations of the $J_1$-$J_2$ model on the A-site spinel
lattice, introducing a certain amount of inversion randomly each time a simulation
run was performed. We also assumed that impurity spins occupying B sites interact with
their nearest neighbors on occupied A sites via a Heisenberg exchange coupling $J_i$.

\begin{figure}[t]
\centering \includegraphics[width=9cm]{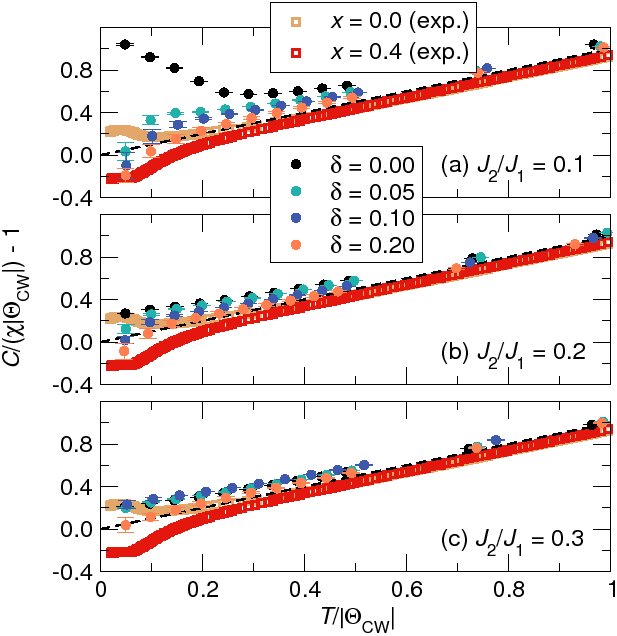}
\caption{(Color online) Magnetic susceptibility results from Monte Carlo simulations plotted in the form of equation \ref{eqn:curie} compared to experimental data. From top to bottom one can see the change in the shape of the low temperature susceptibility curves for increasing competition between $J_2$ and $J_1$. Within each panel the effect of atomic disorder is shown.  Note that $J_i/J_1$ is held constant at 1.7 for all of these simulations. The dotted line represents the expected result for a perfect Curie-Weiss system.}
\label{fig:mc_scans}
\end{figure}

In Fig.\ \ref{fig:mc_scans}, we show the results of a scan of various parameters.  $J_2/J_1$ 
was set to $0.1$, $0.2$, and $0.3$ while the inversion  $\delta$ set to $0.0$, $0.05$, 
$0.10$ and $0.20$. The ratio $J_i/J_1$ was fixed to $1.7$ throughout the scan. 
After plotting the resulting inverse susceptibilities in the manner described in eqn.\ \ref{eqn:curie}, 
we can make a few qualitative observations. As expected, increasing $J_2/J_1$ 
increases the frustration of the system in the sense that it lowers the temperature at which
evidence of ordering appears. Additionally upon introducing even a small amount of inversion, the
behavior of the susceptibility below the Curie-Weiss temperature rapidly changes from 
antiferromagnetic (sharp upward kink) to ferrimagnetic (smooth downturn) in rough 
qualitative agreement with the experimental results. 

We also attempted to use the model to fit the individual susceptibility curves of the
various experimental samples with mixed success. We first conducted simulations of the model with no inversion, 
in which case the fit was controlled by only one parameter, namely $J_2/J_1$. These simulations
were all performed on a system consisting of a cube of 64 conventional unit cells (512 spins), though
for certain values of $J_2/J_1$, simulations were also done using a system of 125 unit cells (1000 spins) to 
check for finite size effects, which were found to be negligible for the susceptibility data. 

\begin{figure}[t]
\centering \includegraphics[width=9cm]{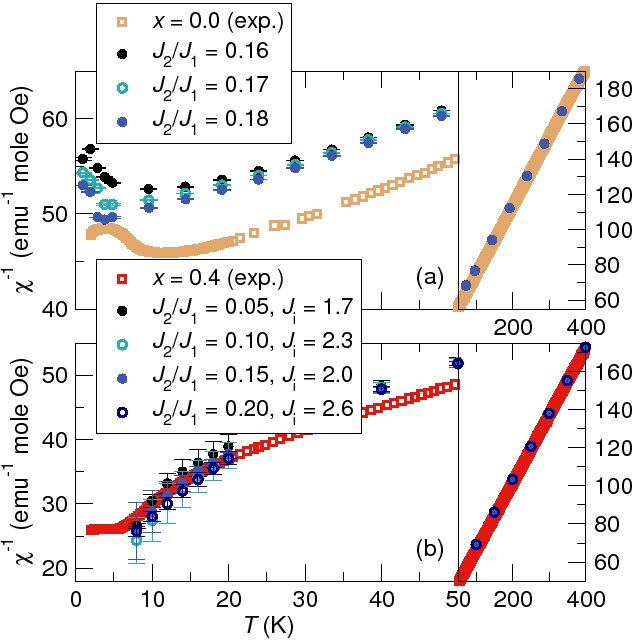}
\caption{(Color online) (a) Susceptibility curves obtained from Monte Carlo simulations where the effect of varying the ratio $J_2/J_1$ is tested in a system with no atomic disorder. (b) Susceptibility curves where the effect of varying the ratio $J_2/J_1$ is tested on a system with a fixed amount of atomic disorder around $\delta=0.36$.}
\label{fig:mc}
\end{figure}

Although all of the inverse susceptibility curves produced from the numerics for the $\delta = 0.0$ 
case were systematically above the experimental data, they showed the correct qualitative behavior 
in that they exhibited an upward turning kink, presumably at the ordering 
temperature $T_N$. By increasing the value of $J_2$ until $J_2/J_1$ is around $0.15$ to $0.18$ (and correspondigly a $J_1$
between $12.1$ and $12.8$ Kelvin), the transition temperature $T_N$ was brought into approximate 
agreement with the experimentally observed value as shown in Fig.\ \ref{fig:mc}. 
We also considered simulations with small amounts of inversion between 1 and 2\% 
given that even pure CoAl$_2$O$_4$ exhibits a small amount of disorder. However, unless we ran our 
simulations with a value of $J_i/J_1$ far from the range that gave a good fit to the 
doped sample with $x=0.4$ (discussed next), such a small amount of disorder changed the 
resulting susceptibilities very little. 

We have also fit the $x=0.4$ data by running simulations with an inversion of $\delta=0.36$. To improve the
disorder averaging, the simulations were done using a $5\times5\times5$ supercell (1000 spin) 
with the data presented here being the average of three independent runs. 
As in the case above, all of the inverse susceptibility data produced by the Monte 
Carlo simulations was systematically larger than the experimental curve excluding the region below the downturn. 
Interestingly the data could be fit equally well by a 
range of $J_2/J_1$ and $J_i/J_1$ values, as long as both parameters were increased together, 
as can be seen from Fig.\ \ref{fig:mc}.
Since it is probably reasonable to assume that $J_2/J_1$ decreases slightly from
the value obtained for the $x=0.0$ sample, we can take $J_2/J_1$ between $0.10$ and $0.15$ 
 and find that the ratio $J_i/J_1$ should lay between $2.0$ and $2.3$ whereas $J_1$ is found to be between 
$8.5$ and $9.7$ K.

It is worth noting at this point that if we judge the quality of the fit in a least-squares sense, the
Curie-Weiss law describes the data better than the numerical susceptibility
data given that the numerical data lies above the experimental data through a wide temperature range 
for which a straight line fits the experimental data almost perfectly. 
Despite this fact, it is valuable to see that the simulation data exhibits the same qualitative low temperature
behavior as the experiments. 
One possible explanation for the deviation between the theory and experimental data, 
aside from the assumption that only nearest and next nearest neighbor exchange play a role, 
is the possible over simplification of the $g$ factor. One way to improve our simulations 
would be to allow for a temperature dependent $g$ or also assigning a different value of $g$ to each site. 

Finally, we attempted to fit the experimental susceptibilities for the samples with $x\geq 0.8$ and were unable to 
obtain a good fit without changing the $J_2/J_1$ and $J_i/J_1$ by an unreasonable amount. For every set of parameters 
considered, the numerically calculated inverse susceptibilities exhibited strong downturns and dropped far below
the experimental curve after initially matching at high temperature. Though the reasons for this 
negative result are somewhat unclear, it seems that either the appropriate 
$J_2/J_1$ and $J_i/J_1$ ratios are very far from those found above or more likely the use of only 
two parameters is an over simplification considering that atomic disorder will 
locally modify exchange pathways and make the true exchange couplings site dependent. This possibility is 
supported by the PDFs presented which show the local bond lengths of the end members is retained upon substitution. 
Such behavior may indicate that substitution actually generates new competing $J$ values which are not accounted for in our 
Monte-Carlo simulations rather than a simple modification to the existing pathways as may have been expected.  

\section{Summary} 

We have attempted to understand the nature of the  magnetic frustration in the A-site
magnetic spinels CoAl$_{2-x}$Ga$_x$O$_4$ by substituting Ga for Al, 
in the hope of decreasing the relative magnetic coupling between near 
and next-near A atom neighbors. We have found, however, that in the compounds presented 
there is a significant mixing of the A and B sites in the Ga-rich samples, 
and this inversion has a significant influence on the magnetic coupling. 
The complexity of the structural changes which occur with substitution of Ga make
isolating the influence of lattice expansion and site mixing a significant challenge.
We have used Monte Carlo calculations to demonstrate the importance of site mixing through a change
of shape in the simulated susceptibility curves which agrees closely with our experimental findings.
By comparing with samples that are perfectly ordered such as Co$_3$O$_4$ and CoRh$_2$O$_4$,
we have demonstrated that there appears to be a trend in which the frustration index $f$ depends weakly 
upon the separation between magnetic ions. We have also used a variety of structural characterizations and magnetic
measurements to demonstrate how both lattice expansion and site mixing can simultaneoulsy influence the frustration parameter.   

\section*{Acknowledgements} We thank A. P. Ramirez for advice in representing
the magnetic data and SungBin Lee for assistance in understanding the Monte Carlo fits. 
We gratefully acknowledge the National Science Foundation for 
support through a Career Award (NSF-DMR\,0449354) to RS, a Graduate Student 
Fellowship to KP, and for MRSEC facilities (Award NSF-DMR\,0520415).
This work has benefited from the use of NPDF at the Lujan Center at Los Alamos 
Neutron Science Center, funded by DOE Office of Basic Energy Sciences under 
DOE contract DE-AC52-06NA25396.

\bibliography{frust}

\begin{thebibliography}{40}
\expandafter\ifx\csname natexlab\endcsname\relax\def\natexlab#1{#1}\fi
\expandafter\ifx\csname bibnamefont\endcsname\relax
  \def\bibnamefont#1{#1}\fi
\expandafter\ifx\csname bibfnamefont\endcsname\relax
  \def\bibfnamefont#1{#1}\fi
\expandafter\ifx\csname citenamefont\endcsname\relax
  \def\citenamefont#1{#1}\fi
\expandafter\ifx\csname url\endcsname\relax
  \def\url#1{\texttt{#1}}\fi
\expandafter\ifx\csname urlprefix\endcsname\relax\def\urlprefix{URL }\fi
\providecommand{\bibinfo}[2]{#2}
\providecommand{\eprint}[2][]{\url{#2}}

\bibitem[{\citenamefont{Ramirez}(1994)}]{GeoFrustRev:1994}
\bibinfo{author}{\bibfnamefont{A.}~\bibnamefont{Ramirez}},
  \bibinfo{journal}{Annu. Rev. Mater. Sci.} \textbf{\bibinfo{volume}{24}},
  \bibinfo{pages}{453} (\bibinfo{year}{1994}).

\bibitem[{\citenamefont{Greedan}(2001)}]{GreedanFrust:2001}
\bibinfo{author}{\bibfnamefont{J.~E.} \bibnamefont{Greedan}},
  \bibinfo{journal}{J. Mater. Chem.} \textbf{\bibinfo{volume}{11}},
  \bibinfo{pages}{37} (\bibinfo{year}{2001}).

\bibitem[{\citenamefont{Fennie and Rabe}(2006)}]{RabeDesign:2006}
\bibinfo{author}{\bibfnamefont{C.~J.} \bibnamefont{Fennie}} \bibnamefont{and}
  \bibinfo{author}{\bibfnamefont{K.~M.} \bibnamefont{Rabe}},
  \bibinfo{journal}{Phys. Rev. Lett.} \textbf{\bibinfo{volume}{97}},
  \bibinfo{pages}{267602} (\bibinfo{year}{2006}).

\bibitem[{\citenamefont{Katsura et~al.}(2005)\citenamefont{Katsura, Nagaosa,
  and Balatsky}}]{Katsura}
\bibinfo{author}{\bibfnamefont{H.}~\bibnamefont{Katsura}},
  \bibinfo{author}{\bibfnamefont{N.}~\bibnamefont{Nagaosa}}, \bibnamefont{and}
  \bibinfo{author}{\bibfnamefont{A.~V.} \bibnamefont{Balatsky}},
  \bibinfo{journal}{Phys. Rev. Lett.} \textbf{\bibinfo{volume}{95}},
  \bibinfo{pages}{057205} (\bibinfo{year}{2005}).

\bibitem[{\citenamefont{Mostovoy}(2006)}]{Mostovoy}
\bibinfo{author}{\bibfnamefont{M.}~\bibnamefont{Mostovoy}},
  \bibinfo{journal}{Phys. Rev. Lett.} \textbf{\bibinfo{volume}{96}},
  \bibinfo{pages}{067601} (\bibinfo{year}{2006}).

\bibitem[{\citenamefont{Sergienko and Dagotto}(2006)}]{Sergienko}
\bibinfo{author}{\bibfnamefont{I.~A.} \bibnamefont{Sergienko}}
  \bibnamefont{and} \bibinfo{author}{\bibfnamefont{E.}~\bibnamefont{Dagotto}},
  \bibinfo{journal}{Phys. Rev. B} \textbf{\bibinfo{volume}{73}},
  \bibinfo{pages}{094434} (\bibinfo{year}{2006}).

\bibitem[{\citenamefont{Tokura}(2006)}]{TokuraScience2006}
\bibinfo{author}{\bibfnamefont{Y.}~\bibnamefont{Tokura}},
  \bibinfo{journal}{Science} \textbf{\bibinfo{volume}{312}},
  \bibinfo{pages}{1481} (\bibinfo{year}{2006}).

\bibitem[{\citenamefont{Yamasaki et~al.}(2006)\citenamefont{Yamasaki, Miyasaka,
  Kaneko, He, Arima, and Tokura}}]{Tokura:2006}
\bibinfo{author}{\bibfnamefont{Y.}~\bibnamefont{Yamasaki}},
  \bibinfo{author}{\bibfnamefont{S.}~\bibnamefont{Miyasaka}},
  \bibinfo{author}{\bibfnamefont{Y.}~\bibnamefont{Kaneko}},
  \bibinfo{author}{\bibfnamefont{J.-P.} \bibnamefont{He}},
  \bibinfo{author}{\bibfnamefont{T.}~\bibnamefont{Arima}}, \bibnamefont{and}
  \bibinfo{author}{\bibfnamefont{Y.}~\bibnamefont{Tokura}},
  \bibinfo{journal}{Phys. Rev. Lett.} \textbf{\bibinfo{volume}{96}},
  \bibinfo{pages}{207204} (\bibinfo{year}{2006}).

\bibitem[{\citenamefont{Lawes et~al.}(2006)\citenamefont{Lawes, Melot, Page,
  Ederer, Hayward, Proffen, and Seshadri}}]{CCO:2006}
\bibinfo{author}{\bibfnamefont{G.}~\bibnamefont{Lawes}},
  \bibinfo{author}{\bibfnamefont{B.}~\bibnamefont{Melot}},
  \bibinfo{author}{\bibfnamefont{K.}~\bibnamefont{Page}},
  \bibinfo{author}{\bibfnamefont{C.}~\bibnamefont{Ederer}},
  \bibinfo{author}{\bibfnamefont{M.~A.} \bibnamefont{Hayward}},
  \bibinfo{author}{\bibfnamefont{T.}~\bibnamefont{Proffen}}, \bibnamefont{and}
  \bibinfo{author}{\bibfnamefont{R.}~\bibnamefont{Seshadri}},
  \bibinfo{journal}{Phys. Rev. B} \textbf{\bibinfo{volume}{74}},
  \bibinfo{pages}{024413} (\bibinfo{year}{2006}).

\bibitem[{\citenamefont{Tackett et~al.}(2007)\citenamefont{Tackett, Lawes,
  Melot, Grossman, Toberer, and Seshadri}}]{MnO:2007}
\bibinfo{author}{\bibfnamefont{R.}~\bibnamefont{Tackett}},
  \bibinfo{author}{\bibfnamefont{G.}~\bibnamefont{Lawes}},
  \bibinfo{author}{\bibfnamefont{B.~C.} \bibnamefont{Melot}},
  \bibinfo{author}{\bibfnamefont{M.}~\bibnamefont{Grossman}},
  \bibinfo{author}{\bibfnamefont{E.~S.} \bibnamefont{Toberer}},
  \bibnamefont{and} \bibinfo{author}{\bibfnamefont{R.}~\bibnamefont{Seshadri}},
  \bibinfo{journal}{Phys. Rev. B} \textbf{\bibinfo{volume}{76}},
  \bibinfo{pages}{024409} (\bibinfo{year}{2007}).

\bibitem[{\citenamefont{Anderson}(1956)}]{BsitePyrochlore:1956}
\bibinfo{author}{\bibfnamefont{P.~W.} \bibnamefont{Anderson}},
  \bibinfo{journal}{Phys. Rev.} \textbf{\bibinfo{volume}{102}},
  \bibinfo{pages}{1008} (\bibinfo{year}{1956}).

\bibitem[{\citenamefont{Bergman et~al.}(2007)\citenamefont{Bergman, Alicea,
  Gull, Trebst, and Balents}}]{BalentsOrderbyDisorder}
\bibinfo{author}{\bibfnamefont{D.}~\bibnamefont{Bergman}},
  \bibinfo{author}{\bibfnamefont{J.}~\bibnamefont{Alicea}},
  \bibinfo{author}{\bibfnamefont{E.}~\bibnamefont{Gull}},
  \bibinfo{author}{\bibfnamefont{S.}~\bibnamefont{Trebst}}, \bibnamefont{and}
  \bibinfo{author}{\bibfnamefont{L.}~\bibnamefont{Balents}},
  \bibinfo{journal}{Nature Physics} \textbf{\bibinfo{volume}{3}},
  \bibinfo{pages}{487} (\bibinfo{year}{2007}).

\bibitem[{\citenamefont{Tristan et~al.}(2005)\citenamefont{Tristan, Hemberger,
  Krimmel, von Nidda, Tsurkan, and Loidl}}]{Tristan:2005}
\bibinfo{author}{\bibfnamefont{N.}~\bibnamefont{Tristan}},
  \bibinfo{author}{\bibfnamefont{J.}~\bibnamefont{Hemberger}},
  \bibinfo{author}{\bibfnamefont{A.}~\bibnamefont{Krimmel}},
  \bibinfo{author}{\bibfnamefont{H.-A.~K.} \bibnamefont{von Nidda}},
  \bibinfo{author}{\bibfnamefont{V.}~\bibnamefont{Tsurkan}}, \bibnamefont{and}
  \bibinfo{author}{\bibfnamefont{A.}~\bibnamefont{Loidl}},
  \bibinfo{journal}{Phys. Rev. B} \textbf{\bibinfo{volume}{72}},
  \bibinfo{pages}{174404} (\bibinfo{year}{2005}).

\bibitem[{\citenamefont{Roth}(1964)}]{RothJPhysFrance}
\bibinfo{author}{\bibfnamefont{W.~L.} \bibnamefont{Roth}}, \bibinfo{journal}{J.
  Phys. France} \textbf{\bibinfo{volume}{25}}, \bibinfo{pages}{507}
  (\bibinfo{year}{1964}).

\bibitem[{\citenamefont{Blasse}(1963)}]{BlasseNewSuperExchange}
\bibinfo{author}{\bibfnamefont{G.}~\bibnamefont{Blasse}},
  \bibinfo{journal}{Philips Res. Rept.} \textbf{\bibinfo{volume}{18}},
  \bibinfo{pages}{383} (\bibinfo{year}{1963}).

\bibitem[{\citenamefont{Tristan et~al.}(2008)\citenamefont{Tristan, Zestrea,
  Behr, Klingeler, B{\"u}chner, von Nidda, Loidl, and Tsurkan}}]{Tristan2008}
\bibinfo{author}{\bibfnamefont{N.}~\bibnamefont{Tristan}},
  \bibinfo{author}{\bibfnamefont{V.}~\bibnamefont{Zestrea}},
  \bibinfo{author}{\bibfnamefont{G.}~\bibnamefont{Behr}},
  \bibinfo{author}{\bibfnamefont{R.}~\bibnamefont{Klingeler}},
  \bibinfo{author}{\bibfnamefont{B.}~\bibnamefont{B{\"u}chner}},
  \bibinfo{author}{\bibfnamefont{H.~A.~K.} \bibnamefont{von Nidda}},
  \bibinfo{author}{\bibfnamefont{A.}~\bibnamefont{Loidl}}, \bibnamefont{and}
  \bibinfo{author}{\bibfnamefont{V.}~\bibnamefont{Tsurkan}},
  \bibinfo{journal}{Phys. Rev. B.} \textbf{\bibinfo{volume}{77}},
  \bibinfo{pages}{094412} (\bibinfo{year}{2008}).

\bibitem[{\citenamefont{Shannon}(1976)}]{Shannon}
\bibinfo{author}{\bibfnamefont{R.~D.} \bibnamefont{Shannon}},
  \bibinfo{journal}{Acta Crystallogr. A} \textbf{\bibinfo{volume}{32}},
  \bibinfo{pages}{751} (\bibinfo{year}{1976}).

\bibitem[{\citenamefont{Porta and Anichini}(1980)}]{CGOcatdistro:1980}
\bibinfo{author}{\bibfnamefont{P.}~\bibnamefont{Porta}} \bibnamefont{and}
  \bibinfo{author}{\bibfnamefont{A.}~\bibnamefont{Anichini}},
  \bibinfo{journal}{J. Chem. Soc. Faraday Trans.}
  \textbf{\bibinfo{volume}{76}}, \bibinfo{pages}{2448} (\bibinfo{year}{1980}).

\bibitem[{\citenamefont{B\'erar and Baldinozzi}(1998)}]{XND}
\bibinfo{author}{\bibfnamefont{J.}~\bibnamefont{B\'erar}} \bibnamefont{and}
  \bibinfo{author}{\bibfnamefont{G.}~\bibnamefont{Baldinozzi}},
  \bibinfo{journal}{IUCr-CPD Newsletter} \textbf{\bibinfo{volume}{20}},
  \bibinfo{pages}{3} (\bibinfo{year}{1998}).

\bibitem[{\citenamefont{Proffen et~al.}(2002)\citenamefont{Proffen, Egami,
  Billinge, Cheetham, Louca, and Parise}}]{NPDF}
\bibinfo{author}{\bibfnamefont{T.}~\bibnamefont{Proffen}},
  \bibinfo{author}{\bibfnamefont{T.}~\bibnamefont{Egami}},
  \bibinfo{author}{\bibfnamefont{S.~J.~L.} \bibnamefont{Billinge}},
  \bibinfo{author}{\bibfnamefont{A.~K.} \bibnamefont{Cheetham}},
  \bibinfo{author}{\bibfnamefont{D.}~\bibnamefont{Louca}}, \bibnamefont{and}
  \bibinfo{author}{\bibfnamefont{J.~B.} \bibnamefont{Parise}},
  \bibinfo{journal}{Appl. Phys. A.} \textbf{\bibinfo{volume}{74}},
  \bibinfo{pages}{S163} (\bibinfo{year}{2002}).

\bibitem[{\citenamefont{Larson and Von~Dreele}(2000)}]{gsas}
\bibinfo{author}{\bibfnamefont{A.~C.} \bibnamefont{Larson}} \bibnamefont{and}
  \bibinfo{author}{\bibfnamefont{R.~B.} \bibnamefont{Von~Dreele}},
  \bibinfo{journal}{Los Alamos National LaboratoryReport LAUR 86-748 (2000)}
  \textbf{\bibinfo{volume}{86}}, \bibinfo{pages}{748} (\bibinfo{year}{2000}).

\bibitem[{\citenamefont{Toby}(2001)}]{expgui}
\bibinfo{author}{\bibfnamefont{B.}~\bibnamefont{Toby}}, \bibinfo{journal}{J.
  Appl. Cryst.} \textbf{\bibinfo{volume}{34}}, \bibinfo{pages}{210}
  (\bibinfo{year}{2001}).

\bibitem[{\citenamefont{Peterson et~al.}(2000)\citenamefont{Peterson, Gutmann,
  Proffen, and Billinge}}]{PDFGETN}
\bibinfo{author}{\bibfnamefont{P.~F.} \bibnamefont{Peterson}},
  \bibinfo{author}{\bibfnamefont{M.}~\bibnamefont{Gutmann}},
  \bibinfo{author}{\bibfnamefont{T.}~\bibnamefont{Proffen}}, \bibnamefont{and}
  \bibinfo{author}{\bibfnamefont{S.~J.~L.} \bibnamefont{Billinge}},
  \bibinfo{journal}{J. Appl. Crystallogr.} \textbf{\bibinfo{volume}{33}},
  \bibinfo{pages}{1192} (\bibinfo{year}{2000}).

\bibitem[{\citenamefont{Farrow et~al.}(2007)\citenamefont{Farrow, Juhas, Liu,
  Bryndin, Bozin, Bloch, Proffen, and Billinge}}]{diffpy}
\bibinfo{author}{\bibfnamefont{C.~L.} \bibnamefont{Farrow}},
  \bibinfo{author}{\bibfnamefont{P.}~\bibnamefont{Juhas}},
  \bibinfo{author}{\bibfnamefont{J.~W.} \bibnamefont{Liu}},
  \bibinfo{author}{\bibfnamefont{D.}~\bibnamefont{Bryndin}},
  \bibinfo{author}{\bibfnamefont{E.~S.} \bibnamefont{Bozin}},
  \bibinfo{author}{\bibfnamefont{J.}~\bibnamefont{Bloch}},
  \bibinfo{author}{\bibfnamefont{T.}~\bibnamefont{Proffen}}, \bibnamefont{and}
  \bibinfo{author}{\bibfnamefont{S.~J.~L.} \bibnamefont{Billinge}},
  \bibinfo{journal}{J. Phys.: Condens. Matter} \textbf{\bibinfo{volume}{19}},
  \bibinfo{pages}{335219 (7pp)} (\bibinfo{year}{2007}).

\bibitem[{\citenamefont{Albuquerque et~al.}(2007)\citenamefont{Albuquerque,
  Alet, Corboz, Dayal, Feiguin, Fuchs, Gamper, Gull, Gurtler, and {\it et
  al.}}}]{ALPS}
\bibinfo{author}{\bibfnamefont{A.}~\bibnamefont{Albuquerque}},
  \bibinfo{author}{\bibfnamefont{F.}~\bibnamefont{Alet}},
  \bibinfo{author}{\bibfnamefont{P.}~\bibnamefont{Corboz}},
  \bibinfo{author}{\bibfnamefont{P.}~\bibnamefont{Dayal}},
  \bibinfo{author}{\bibfnamefont{A.}~\bibnamefont{Feiguin}},
  \bibinfo{author}{\bibfnamefont{S.}~\bibnamefont{Fuchs}},
  \bibinfo{author}{\bibfnamefont{L.}~\bibnamefont{Gamper}},
  \bibinfo{author}{\bibfnamefont{E.}~\bibnamefont{Gull}},
  \bibinfo{author}{\bibfnamefont{S.}~\bibnamefont{Gurtler}}, \bibnamefont{and}
  \bibinfo{author}{\bibfnamefont{A.~H.} \bibnamefont{{\it et al.}}},
  \bibinfo{journal}{J.\ Magn.\ Magn.\ Mater.} \textbf{\bibinfo{volume}{310}},
  \bibinfo{pages}{1187} (\bibinfo{year}{2007}).

\bibitem[{\citenamefont{Joubert et~al.}(1998)\citenamefont{Joubert, Cern{\'y},
  Latroche, Percheron-Gu{\'e}gan, and Yvon}}]{Joubert}
\bibinfo{author}{\bibfnamefont{J.-M.} \bibnamefont{Joubert}},
  \bibinfo{author}{\bibfnamefont{R.}~\bibnamefont{Cern{\'y}}},
  \bibinfo{author}{\bibfnamefont{M.}~\bibnamefont{Latroche}},
  \bibinfo{author}{\bibfnamefont{A.}~\bibnamefont{Percheron-Gu{\'e}gan}},
  \bibnamefont{and} \bibinfo{author}{\bibfnamefont{K.}~\bibnamefont{Yvon}},
  \bibinfo{journal}{J. Appl. Crystallogr.} \textbf{\bibinfo{volume}{31}},
  \bibinfo{pages}{327} (\bibinfo{year}{1998}).

\bibitem[{\citenamefont{Miller}(1959)}]{Miller1959}
\bibinfo{author}{\bibfnamefont{A.}~\bibnamefont{Miller}}, \bibinfo{journal}{J.
  Appl. Phys.} \textbf{\bibinfo{volume}{30}}, \bibinfo{pages}{24S}
  (\bibinfo{year}{1959}).

\bibitem[{\citenamefont{Nakatsuka et~al.}(2003)\citenamefont{Nakatsuka, Ikeda,
  Yamasaki, Nakayama, and Mizota}}]{nakatsuka_cation_2003}
\bibinfo{author}{\bibfnamefont{A.}~\bibnamefont{Nakatsuka}},
  \bibinfo{author}{\bibfnamefont{Y.}~\bibnamefont{Ikeda}},
  \bibinfo{author}{\bibfnamefont{Y.}~\bibnamefont{Yamasaki}},
  \bibinfo{author}{\bibfnamefont{N.}~\bibnamefont{Nakayama}}, \bibnamefont{and}
  \bibinfo{author}{\bibfnamefont{T.}~\bibnamefont{Mizota}},
  \bibinfo{journal}{Solid State Communications} \textbf{\bibinfo{volume}{128}},
  \bibinfo{pages}{85} (\bibinfo{year}{2003}).

\bibitem[{\citenamefont{Nakatsuka et~al.}(2006)\citenamefont{Nakatsuka, Ikeda,
  Nakayama, and Mizota}}]{Nakatsuka:wm2007}
\bibinfo{author}{\bibfnamefont{A.}~\bibnamefont{Nakatsuka}},
  \bibinfo{author}{\bibfnamefont{Y.}~\bibnamefont{Ikeda}},
  \bibinfo{author}{\bibfnamefont{N.}~\bibnamefont{Nakayama}}, \bibnamefont{and}
  \bibinfo{author}{\bibfnamefont{T.}~\bibnamefont{Mizota}},
  \bibinfo{journal}{Acta Crystallographica Section E}
  \textbf{\bibinfo{volume}{62}}, \bibinfo{pages}{i109} (\bibinfo{year}{2006}).

\bibitem[{\citenamefont{Hill et~al.}(1979)\citenamefont{Hill, Craig, and
  Gibbs}}]{Hill1979}
\bibinfo{author}{\bibfnamefont{R.~J.} \bibnamefont{Hill}},
  \bibinfo{author}{\bibfnamefont{J.~R.} \bibnamefont{Craig}}, \bibnamefont{and}
  \bibinfo{author}{\bibfnamefont{G.~V.} \bibnamefont{Gibbs}},
  \bibinfo{journal}{Phys. Chem. Minerals} \textbf{\bibinfo{volume}{4}},
  \bibinfo{pages}{317} (\bibinfo{year}{1979}).

\bibitem[{\citenamefont{Jeong et~al.}(2001)\citenamefont{Jeong,
  Mohiuddin-Jacobs, Petkov, Billinge, and Kycia}}]{JeongPRB}
\bibinfo{author}{\bibfnamefont{I.-K.} \bibnamefont{Jeong}},
  \bibinfo{author}{\bibfnamefont{F.}~\bibnamefont{Mohiuddin-Jacobs}},
  \bibinfo{author}{\bibfnamefont{V.}~\bibnamefont{Petkov}},
  \bibinfo{author}{\bibfnamefont{S.~J.~L.} \bibnamefont{Billinge}},
  \bibnamefont{and} \bibinfo{author}{\bibfnamefont{S.}~\bibnamefont{Kycia}},
  \bibinfo{journal}{Phys. Rev. B} \textbf{\bibinfo{volume}{63}},
  \bibinfo{pages}{205202} (\bibinfo{year}{2001}).

\bibitem[{\citenamefont{Fiorani and Viticoli}(1978)}]{Fiorani1978}
\bibinfo{author}{\bibfnamefont{D.}~\bibnamefont{Fiorani}} \bibnamefont{and}
  \bibinfo{author}{\bibfnamefont{S.}~\bibnamefont{Viticoli}},
  \bibinfo{journal}{Sol. State Comm.} \textbf{\bibinfo{volume}{25}},
  \bibinfo{pages}{155} (\bibinfo{year}{1978}).

\bibitem[{\citenamefont{Suzuki et~al.}(2007)\citenamefont{Suzuki, Nagai,
  Nohara, and Takagi}}]{SuzukiCoAl2O4}
\bibinfo{author}{\bibfnamefont{T.}~\bibnamefont{Suzuki}},
  \bibinfo{author}{\bibfnamefont{H.}~\bibnamefont{Nagai}},
  \bibinfo{author}{\bibfnamefont{M.}~\bibnamefont{Nohara}}, \bibnamefont{and}
  \bibinfo{author}{\bibfnamefont{H.}~\bibnamefont{Takagi}},
  \bibinfo{journal}{J. Phys.: Condens. Matter.} \textbf{\bibinfo{volume}{19}},
  \bibinfo{pages}{145265} (\bibinfo{year}{2007}).

\bibitem[{\citenamefont{Melot et~al.}(2009)\citenamefont{Melot, Drewes,
  Seshadri, Stoudenmire, and Ramirez}}]{Melot2009}
\bibinfo{author}{\bibfnamefont{B.~C.} \bibnamefont{Melot}},
  \bibinfo{author}{\bibfnamefont{J.~E.} \bibnamefont{Drewes}},
  \bibinfo{author}{\bibfnamefont{R.}~\bibnamefont{Seshadri}},
  \bibinfo{author}{\bibfnamefont{E.~M.} \bibnamefont{Stoudenmire}},
  \bibnamefont{and} \bibinfo{author}{\bibfnamefont{A.~P.}
  \bibnamefont{Ramirez}}, \bibinfo{journal}{J. Phys.: Condens. Matter}
  \textbf{\bibinfo{volume}{21}}, \bibinfo{pages}{216007}
  (\bibinfo{year}{2009}).

\bibitem[{\citenamefont{Day and Selbin}(1960)}]{DaySelbin}
\bibinfo{author}{\bibfnamefont{M.~C.} \bibnamefont{Day}} \bibnamefont{and}
  \bibinfo{author}{\bibfnamefont{J.}~\bibnamefont{Selbin}},
  \emph{\bibinfo{title}{Theoretical Inorganic Chemistry}}
  (\bibinfo{publisher}{Reinhold Book Corporation}, \bibinfo{year}{1960}),
  \bibinfo{edition}{2nd} ed.

\bibitem[{\citenamefont{Cossee and van
  Arkel}(1960)}]{Cossee_JPhysChemSolids1960}
\bibinfo{author}{\bibfnamefont{P.}~\bibnamefont{Cossee}} \bibnamefont{and}
  \bibinfo{author}{\bibfnamefont{A.~E.} \bibnamefont{van Arkel}},
  \bibinfo{journal}{J. Phys. Chem. Solid.} \textbf{\bibinfo{volume}{15}},
  \bibinfo{pages}{1} (\bibinfo{year}{1960}).

\bibitem[{\citenamefont{Ederer and Komelj}(2007)}]{ederer:064409}
\bibinfo{author}{\bibfnamefont{C.}~\bibnamefont{Ederer}} \bibnamefont{and}
  \bibinfo{author}{\bibfnamefont{M.}~\bibnamefont{Komelj}},
  \bibinfo{journal}{Physical Review B (Condensed Matter and Materials Physics)}
  \textbf{\bibinfo{volume}{76}}, \bibinfo{eid}{064409}
  (pages~\bibinfo{numpages}{9}) (\bibinfo{year}{2007}).

\bibitem[{\citenamefont{Soubeyroux et~al.}(1986)\citenamefont{Soubeyroux,
  Fiorani, and Agostinelli}}]{SpinGlassCGO:1986}
\bibinfo{author}{\bibfnamefont{J.}~\bibnamefont{Soubeyroux}},
  \bibinfo{author}{\bibfnamefont{D.}~\bibnamefont{Fiorani}}, \bibnamefont{and}
  \bibinfo{author}{\bibfnamefont{E.}~\bibnamefont{Agostinelli}},
  \bibinfo{journal}{J. Magn. Magn. Mater.} \textbf{\bibinfo{volume}{54-57}},
  \bibinfo{pages}{83} (\bibinfo{year}{1986}).

\bibitem[{\citenamefont{Diaz et~al.}(2006)\citenamefont{Diaz, de~Brion,
  Chouteau, Canals, Simonet, and Strobel}}]{Diaz2006}
\bibinfo{author}{\bibfnamefont{S.}~\bibnamefont{Diaz}},
  \bibinfo{author}{\bibfnamefont{S.}~\bibnamefont{de~Brion}},
  \bibinfo{author}{\bibfnamefont{G.}~\bibnamefont{Chouteau}},
  \bibinfo{author}{\bibfnamefont{B.}~\bibnamefont{Canals}},
  \bibinfo{author}{\bibfnamefont{V.}~\bibnamefont{Simonet}}, \bibnamefont{and}
  \bibinfo{author}{\bibfnamefont{P.}~\bibnamefont{Strobel}},
  \bibinfo{journal}{Phys. Rev. B} \textbf{\bibinfo{volume}{74}},
  \bibinfo{pages}{092404} (\bibinfo{year}{2006}).

\bibitem[{\citenamefont{Binder and Young}(1986)}]{Binder1986}
\bibinfo{author}{\bibfnamefont{K.}~\bibnamefont{Binder}} \bibnamefont{and}
  \bibinfo{author}{\bibfnamefont{A.~P.} \bibnamefont{Young}},
  \bibinfo{journal}{Rev. Mod. Phys.} \textbf{\bibinfo{volume}{58}},
  \bibinfo{pages}{801} (\bibinfo{year}{1986}).

\end{thebibliography}

\clearpage

\end{document}